\documentclass[amsmath,amssymb, aps, pra,reprint,footinbib,floatfix,superscriptaddress]{revtex4-2}

\usepackage{float}
\usepackage{amsmath,amssymb}
\usepackage[english]{babel}
\usepackage{upgreek}
\usepackage{dsfont}
\usepackage{graphicx}
\usepackage{color}
\usepackage{bbold}
\usepackage{hyperref}
\usepackage[authormarkup=none]{changes}

\begin{document}
\title{Photon Bound States in Coupled Waveguides}

\author{Björn Schrinski}
\affiliation{Center for Hybrid Quantum Networks (Hy-Q), The Niels Bohr Institute, University of Copenhagen, Blegdamsvej 17, 2100 Copenhagen, Denmark}
\author{Johan A. Brimer}
\affiliation{Center for Hybrid Quantum Networks (Hy-Q), The Niels Bohr Institute, University of Copenhagen, Blegdamsvej 17, 2100 Copenhagen, Denmark}
\author{Anders S. S\o rensen}
\affiliation{Center for Hybrid Quantum Networks (Hy-Q), The Niels Bohr Institute, University of Copenhagen, Blegdamsvej 17, 2100 Copenhagen, Denmark}

\begin{abstract}
Photon bound states have been identified as particular solutions to the scattering of two photons from a single emitter, but from these 
results the full nature of these states remains elusive. We study a novel, clear and unambiguous signature that these bound states are truly bound. To this end 
we consider a new configuration of close-by waveguides, each chirally coupled to two-level emitters. We show that   in this system the photon bound states behave like rigid molecules, where photons do not tunnel  individually but rather collectively, such that there is rarely a single photon in each waveguide. We further identify new classes of bound states in this system.

\end{abstract}

\maketitle

\emph{Introduction--} 
Photons hardly interact directly, but can interact indirectly through coupling to matter \cite{chang2014quantum}. In systems with photons coupled to quantum emitters it makes intuitive sense that effects like the  exclusion of multiple excitations of the same emitter combined with stimulated emission lead to strong correlations in the emitted photons. These correlations can reach the degree that photons are ``bound" to each other \cite{shen2007strongly1,shen2007strongly2,firstenberg2013attractive,liang2018observation}, interact attractively or repulsively \cite{bakkensen2021photonic,calajo2022emergence,iversen2021strongly},
scatter inelasically \cite{ke2019inelastic,schrinski2022polariton}
and may even form some kind of quasi matter. 
In this way, setups consisting of waveguides coupled to two- or multi-level emitters \cite{sheremet2023waveguide} can be used to create 
a range of interesting phenomena, e.g. from new quantum many-body states of light \cite{hartmann2007strong,chang2008crystallization,kiffner2010dissipation,kiffner2011dissipation,hafezi2013non,kapit2014induced,albrecht2019subradiant,zhang2019theory,zhang2022free,schrinski2022polariton} to 
 non-linear phase shifts enabling quantum computation schemes \cite{brod2016passive,brod2016two,schrinski2022passive,shapourian2023modular}.

Here we consider bound states of photons. Such ``photonic-molecules" are the quantum limit of optical solitons \cite{lai1989quantum1,lai1989quantum2,mahmoodian2020dynamics}. They have been predicted in a range of systems such as two- \cite{yudson1985dynamics,shen2007strongly1,shen2007strongly2,zhang2020subradiant,bakkensen2021photonic,calajo2022emergence} and three-level systems \cite{iversen2021strongly}
coupled to waveguides as well as  atomic gases with Rydberg interaction \cite{kiffner2013three,bienias2014scattering,letscher2018mobile}. Experimentally, bound states have been observed in Rydberg atomic ensembles \cite{liang2018observation} and with a quantum dot coupled to an optical cavity \cite{tomm2023photon}.
The dispersion relation of the bound states can be related  to particles possessing mass, but such  analogies raise the question of whether this is just a mathematical analogy  or if the bound photons also act in a collective manner when being manipulated, i.e.\,do they behave as ``truly distinct physical objects" \cite{mahmoodian2020dynamics}? 

\begin{figure}
  \centering
  \includegraphics[width=0.48\textwidth]{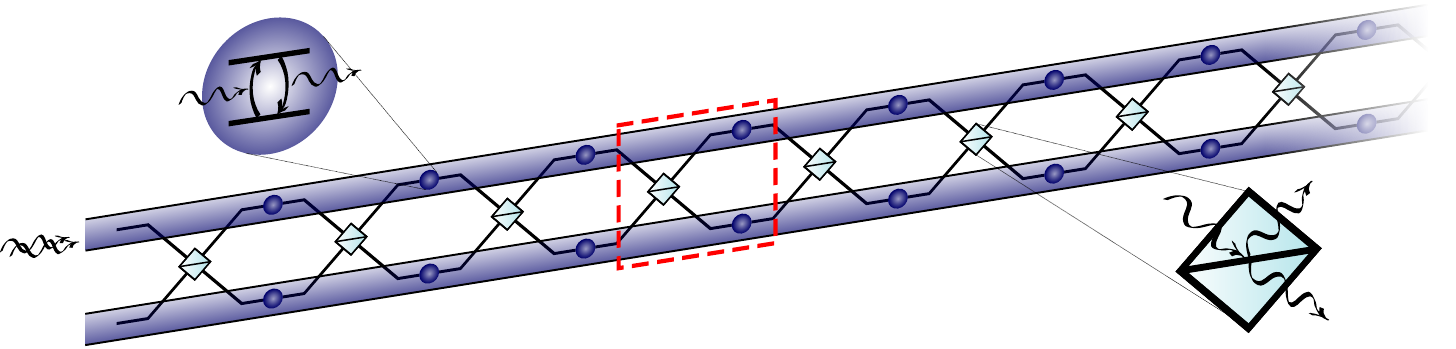}
  \caption{The considered setup  consists of two one-dimensional photonic waveguides each chirally coupled to two-level quantum emitters. The waveguides are close enough to allow tunneling of photons from one to the other, which is modelled as beamsplitters with a high reflection probability. The red rhomb marks the unit cell 
  used to describe the propagation. 
  }
\label{fig:BeamsplitterSetup}
\end{figure}

We want to continue this train of thought  by considering the very basic question, do photons in a bound state stick together when we try to separate them? To answer this, we consider
 a scenario where we allow photons to tunnel between two nearby waveguides, each chirally coupled to two-level emitters that  decay by emitting photons \cite{lodahl2017chiral,hauff2022chiral} into the waveguide with a decay rate $\Gamma$. Such chiral systems, where emitters only emit photons in a single direction have been experimentally realized in Refs. \cite{sollner2015deterministic,le2015nanophotonic,lodahl2017chiral,barik2020chiral,hauff2022chiral}.  For a single waveguide, this configuration is known to  allow for the creation of photon bound states  \cite{mahmoodian2020dynamics,iversen2021strongly}. We extend the model to  two coupled waveguides to investigate what happens to the bound states when the photons can tunnel between the waveguides and thereby  break apart the  bound state. Coherent tunneling is described by 
inserting a beamsplitter transformation after each emitter with a very large reflection probability $P\simeq 1-\theta^2$ where $\theta\ll 1$, see Fig.\,\ref{fig:BeamsplitterSetup}. In the linear scenario, i.e. without photon interactions, 
the individual photons would independently switch from one waveguide to the other and back again. For truly bound photons this slow continuous tunnelling should occur in pairs and should  be suppressed as the probability to tunnel scales as $(1-P)^2\simeq \theta^4$ since the photons have to switch waveguides simultaneously.  In this paper we show  that this is indeed the case, how new classes of bound states emerge in the setup, and that the different classes of bound and free photons 
can be distinguished by their deviating group velocities.    

\emph{Bound states--} 
The description of non-linear scattering of photons coupled to a single two-level system in general, and the emerging bound states in particular, was pioneered by Yudson \cite{yudson1985dynamics} and finalized by Shen and Fan \cite{shen2007strongly1,shen2007strongly2} showing that the two photon Hilbert space can be divided into ``free" and ``bound" subspaces, described by the eigenstates  $|W_{k_1,k_2}\rangle$ and $|B_K\rangle$, respectively, where $K=k_1+k_2$ is the total energy (for simplicity we set group velocity and $\hbar$ equal to unity such that decay rates and wavenumbers have the same dimensions). The bound states have the characteristic feature of being localized in position space \cite{shen2007strongly2}, with the wavefunction 
\begin{align}\label{eq:boundstate}
\langle x_1,x_2|B_K\rangle = \frac{\sqrt{\Gamma}}{\sqrt{4\pi}} 
e^{i K X-\Gamma|\Delta|/2},
\end{align}
being exponentially suppressed with  the relative distance coordinate $\Delta=x_1-x_2$  for two photons at locations $x_1$ and $x_2$ with center of mass coordinate $X=(x_1+x_2)/2$. Here $\Gamma$ denotes the decay rate of the emitter into the waveguide and we assume for simplicity that there are no other decay channels. On the other hand the unbound states are described by \cite{shen2007strongly2}
\begin{align}
\langle x_1,x_2|W_{k_1,k_2}\rangle =& \frac{1}{\sqrt{2\pi^2}} 
e^{i K X}\nonumber\\
&\times[2q\cos(q \Delta)-\Gamma \mathrm{sgn}(x)\sin(q \Delta)],
\end{align}
where $q=(k_1-k_2)/2$ is the relative momentum.

The scattering process of a single emitter can  be expressed in the complete basis 
of bound and unbound states via the (non-linear) scattering matrix \cite{shen2007strongly2}
\begin{align}
\mathrm{S}=\sum_{k_1\leq k_2}t_{k_1} t_{k_2} |W_{{k_1},{k_2}}\rangle\langle W_{{k_1},{k_2}}|+
\sum_K T_K |B_K\rangle \langle B_K|,
\end{align}
where 
\begin{align}
t_k = \frac{k-i\Gamma/2}{k+i\Gamma/2},\quad\mathrm{and}\quad T_K = \frac{K-2i\Gamma}{K+2i\Gamma}        
\end{align}
are the linear phase shift acquired  for an individual photon and the  non-linear phase shift of the bound state, respectively. 
For chiral systems this dynamics can by extended to multiple emitters by repeated application of the scattering matrix. Such an extension to more emitters and photons opens up rich physics \cite{ramos2014quantum,mirza2016multiqubit,sanchez2020chiral,mahmoodian2020dynamics,lorenzo2021intermittent,iversen2022self} with e.g. bound states containing larger numbers of photons with associated  increasing group velocities. The resulting spatial separation upon propagation allows for separating the different bound states and for example to study scattering with other photon states \cite{mahmoodian2020dynamics}. 

\emph{Mathematical description--} We wish to describe a system where photons are distributed in two neighbouring waveguides denoted by $a$ and $b$. In the momentum basis, states containing two photons can be written as 
\begin{align}
|\psi\rangle=\sum_{m,n\in{a,b}}\int \mathrm{d}K \mathrm{d}q\,
\psi_{mn}(K,q)|K,q\rangle_{mn}
\end{align}
with basis states 
\begin{align}
|K,q\rangle_{mn}=\mathsf{c}_m^\dagger(K/2-q)\mathsf{c}_n^\dagger(K/2+q)|\mathrm{vac}\rangle,
\end{align}
where $\mathsf{c}_a^\dagger(k)$ or $\mathsf{c}_b^\dagger(k)$ create a photon with momentum $k$ in the upper or lower mode, respectively.
We want to describe the system depicted in Fig.\,\ref{fig:BeamsplitterSetup} by merging the non-linear scattering and the tunneling transformation into one combined transformation $\mathrm{T}$ for each unit cell as defined by the red rhomb. We can project the different combinations of the two photons in the two waveguides into  a vector 
$\mathbf{v}=|\psi_{aa},\psi_{ab},\psi_{ba},\psi_{bb}\rangle$ with $|\psi_{mn}\rangle=\int \mathrm{d}K \mathrm{d}q\,
|K,q\rangle_{mn}\langle K,q|_{mn}|\psi\rangle$. For bosonic exchange symmetry reasons the second and third entry of $\mathbf{v}$ are  always equal.
In this vector notation each unit cell transformation takes on the form
\begin{align}\label{eq:unitTransformation}
\mathrm{T}=\mathrm{diag}(\mathrm{S},\mathrm{S}_\mathrm{lin},\mathrm{S}_\mathrm{lin},\mathrm{S})(\mathrm{M}\otimes\mathrm{M}),
\end{align}
with the unitary tunneling matrix 
\begin{align}
\mathrm{M}=
\begin{pmatrix}
\cos(\theta)&i\sin(\theta)\\
i\sin(\theta)&\cos(\theta)
\end{pmatrix}
\end{align}
and the linear scattering matrix
\begin{align}
\mathrm{S}_\mathrm{lin}=\sum_{m,n\in{a,b}}\sum_{K,q}t_{K/2-q} t_{K/2+q} |K,q\rangle_{mn}\langle K,q|_{mn}.
\end{align}
Here, the linear scattering describes the evolution of photons in separate waveguides, where there is no non-linear interaction between the photons.  
The state after $N$ scattering events then reads $\mathrm{T}^N\mathbf{v}$. 

\begin{figure}
  \centering
  \includegraphics[width=0.49\textwidth]{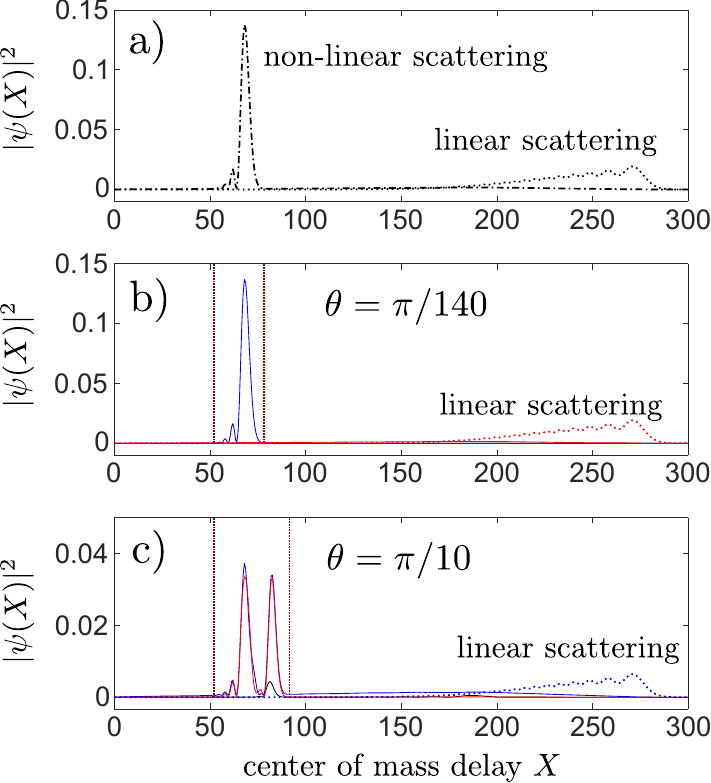}
  \caption{Center of mass  distribution  after $N=70$ scatterings. a) For a single waveguide in the linear regime (dotted curve) photons on resonance move the slowest with Wigner delay $4N/\Gamma$ and strong dispersion due to the frequency width of the input state. Turning on the non-linearity (dashed dotted line) produces a bound state with group delay $N/\Gamma$ and much less dispersion. b) A small tunneling probability ($\theta=\pi/140$) between the two waveguides only allows  unbound photons to switch  to the $bb$ mode (red curve) whereas the bound state remains practically in its entirety in the $aa$ mode (blue curve). The mode with one photon in each path $ab$ (black curve, hardly visible) is unoccupied. c) Increasing the tunneling amplitude ($\theta=\pi/10$) shows the emergence of two classes of bound states with different group velocities, colors as in b). }
\label{fig:SpatialMarginal}
\end{figure}

\emph{Dynamics of bound states in coupled waveguides--} 
We now investigate various scenarios of photons scattering for the two waveguide setup. We consider a Gaussian product state of two photons entering the upper waveguide, i.e.\,$\psi_{mn}(K,q)\propto\delta_{ma}\delta_{na}e^{-(k_1^2+k_2^2)/2\sigma_k^2}$, with $\sigma_k=\Gamma/3$ to maximize the amplitude of the bound state \cite{mahmoodian2020dynamics}. The center of mass marginal population of photons after the last (here 70th) emitter  are shown in Fig.~\ref{fig:SpatialMarginal} after Fourier transforming to position space and tracing over the relative coordinate $\Delta$. As a reference we first consider a single waveguide in Fig.~\ref{fig:SpatialMarginal} a). As discussed in detail in Ref.~\cite{mahmoodian2020dynamics} there is  a stark contrast between the non-linear and linear cases, where we replace $\mathrm{S}$ by $\mathrm{S}_\mathrm{lin}$. The combination of saturation of the emitters and stimulated emission between the photons leads to a faster group velocity for the bound state. As a consequence, when considering the Wigner delay, i.e. the delay relative to a non-interacting pulse, the linear case  results in a slow, strongly dispersed state delayed by $4N/\Gamma$ whereas the bound state traverses the setup much faster with a delay $N/\Gamma$.

Allowing for tunneling with a finite $\theta$ in the linear setup, the population completely shifts from mode $a$ to $b$  after $\pi/2\theta$ scattering events as shown in Fig.~\ref{fig:SpatialMarginal} b).  After an integer times $\pi/\theta$ scatterings, the population is back in the original waveguide, c.f.  Fig.~\ref{fig:SpatialMarginal} c). 
In contrast, for the non-linear case we expect the emerging bound state to remain in the same waveguide for $\theta\ll 1$. This is confirmed in Fig.~\ref{fig:SpatialMarginal} b) where we observe that the bound state population remains almost unaffected by the tunneling, whereas the unbound population has tunnelled  to the other waveguide.

\begin{figure}
  \centering
  \includegraphics[width=0.49\textwidth]{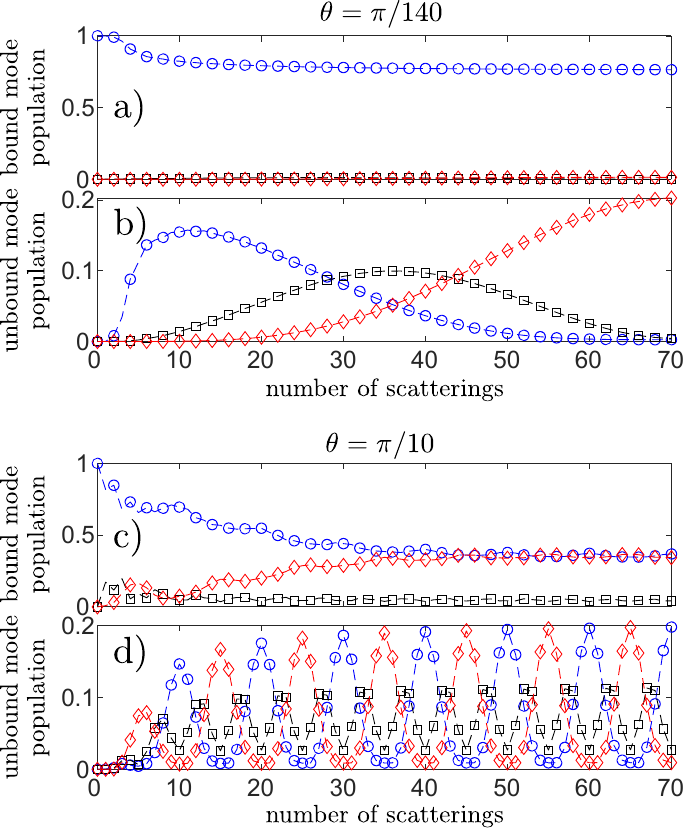}
  \caption{Populations of bound and unbound photons in the $aa$ mode (blue circles), $ab$ mode (black squares), and $bb$ mode (red diamonds) after each scattering for two different tunnel probabilities, $\theta=\pi/140$ (a,b) and $\theta=\pi/10$ (c,d). To distinguish the bound (a,c) and unbound (b,d) populations we cut out windows around the bound states in the center of mass marginals, see Fig.~\ref{fig:SpatialMarginal}. Since we observe almost no dispersion and a group delay $\propto N$ \cite{mahmoodian2020dynamics} we scale the boundaries of the windows with $N$. Outside the windows the photons oscillate individually between the two waveguides with intermediate population of the $ab$ mode. Inside the window the bound photons only jump in pairs. 
  The bound state windows 
  include a residue of unbound photons, leading to small oscillations in  the bound state populations. In (a,b,c) we have removed every second  point for better readability. The dashed lines are guides to the eye.}
\label{fig:Populations}
\end{figure}

To further investigate the tunnelling dynamics we divide the outgoing population into bound and unbound states based on their delay as indicated by the rectangle in Fig.~\ref{fig:SpatialMarginal} b). In  Fig.~\ref{fig:Populations} a) and b), we then consider  how the population is distributed on the two waveguides inside and outside this window when we vary the number of scatterings. 
The window is defined such that all population is inside the window at $N=0$ (translating to $X\in[-6.0,6.0]$) and grows  linearly with $N$ to reach the size indicated in Fig.~\ref{fig:SpatialMarginal} b) (translating to $X\in[53.3,80.0]$). For small $N$ the bound and unbound populations overlap and the population inside the window is unity in Fig.~\ref{fig:Populations} a). As the bound and unbound populations separate, the population inside the window reaches a population around  $\sim 78\%$, which is the bound state population for the  chosen input state with $\sigma_k=\Gamma/3$. Furthermore, for the bound state both photons  remain in the initial waveguide and do not separate. However, in case of the unbound contributions, Fig.~\ref{fig:Populations} b), the photons tunnel independently and go to the other waveguide via an intermediate state with one photon in each waveguide. 
This behavior confirms the compound molecule-like behaviour of the photon bound state.

 An even more convincing demonstration of the inseparable nature of bound photons would be if  the bound photons  collectively switch modes. To show this,  we increase the tunneling probability to $\theta=\pi/10$ in Fig.~\ref{fig:Populations} c) and d) so that free photons individually switch back and forth between the modes seven times after traversing  $N=70$ emitters. The bound state instead slowly tunnels directly from mode $aa$ to $bb$ until a equilibrium  is reached where it is equally likely to find the bound state in one or the other mode.  Importantly, we again only find a small probability for the bound photons to be in two different waveguides. 
 Here we chose the windows to be  $X\in[-6.0,6.0]$ at $N=0$ shifting again linearly to $X\in[53.3,90,6]$ at $N=70$ as shown in Fig.~\ref{fig:SpatialMarginal} c).

Due to the rather appreciable tunnelling $\theta=\pi/10$, the spatial distribution 
 differs significantly from the case without tunnelling. 
We now observe  two distinct bound states of $\mathrm{T}$, which travel at different group velocities, as shown in Fig.~\ref{fig:SpatialMarginal} c). The first  is the anti-symmetric superposition of bound states in each waveguide $|B_K\rangle_{aa}-|B_K\rangle_{bb}$. Since this state is antisymmetric under the exchange of $a$ and $b$ it does not couple to $ab$ components by tunnelling and it is an exact eigenstate for all $\theta$. Furthermore it has the same eigenvalue as the single waveguide bound state and thus travels at the same group velocity. The symmetric superposition $|B_K\rangle_{aa}+|B_K\rangle_{bb}$, however, is only an eigenstate if $\theta=0$. For finite $\theta$ it will couple to components with photons in different modes and as seen in Fig.~\ref{fig:SpatialMarginal} c) this slightly decreases the group velocity as photons in separate waveguides behave as free photons and thus move slower. We will explore these new bound states next.

\begin{figure}
  \centering
  \includegraphics[width=0.49\textwidth]{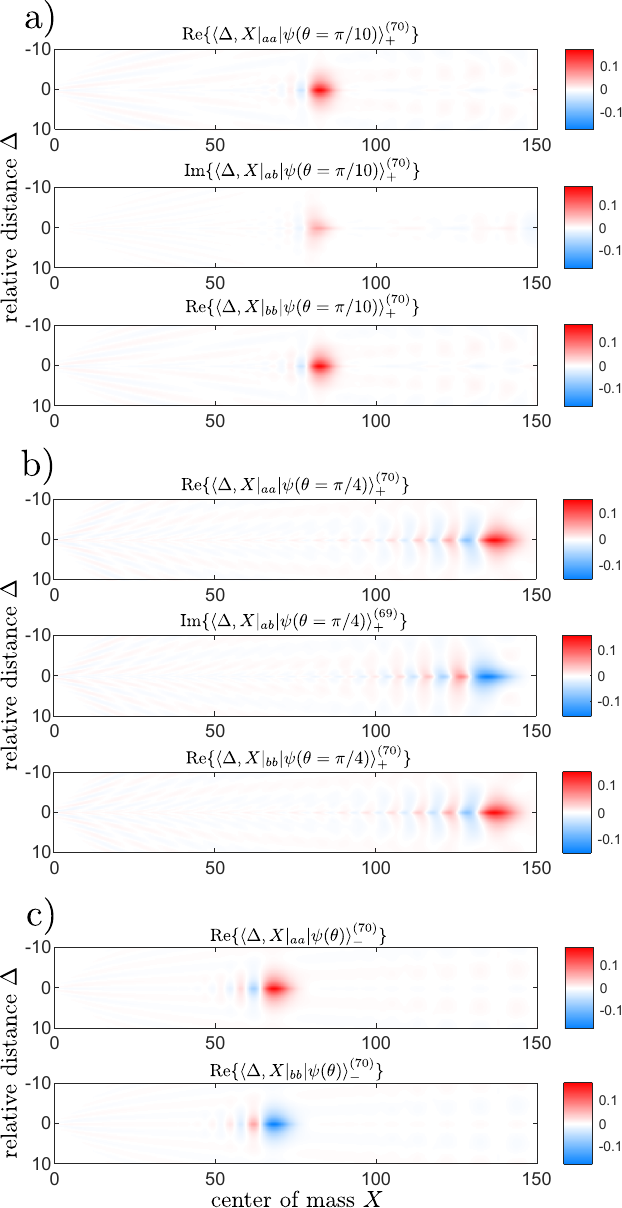}
   \caption{Real (or imaginary) part of the spatial wave functions with both photons in the upper ($aa$), lower ($bb$), or one in each path ($ab$). The respective output state $|\psi(\theta)\rangle_\pm^{(N)}$ after $N$ scatterings is achieved with either a symmetric (+) or anti-symmetric (-) input Gaussian product state (see main text). a) For finite tunneling amplitudes $\theta=\pi/10$ the symmetric  bound state 
   attains a small contribution in the $ab$ mode. c) For $\theta=\pi/4$, the Hong-Ou-Mandel effect lets the  bound state flip after every scattering between both photons being in the same path and both photons being in different paths. Note that the central plot is for $N=69$ and the top and lower plots are for $N=70$. c) For reference, the anti-symmetric bound state is a stable eigenstate for all $\theta$. In all scenarios leaking from the bound states emerges as a faint leaking pattern over almost all phase space because the bound and unbound contribution cannot be distinguished exactly for  finite number of scatterings.}
\label{fig:Wavefunctions}
\end{figure}

\emph{Stability of bound states--}
To study these new classes of bound states in isolation,  we chose as input state a Gaussian product state in a (anti)symmetric superposition $\psi_{mn}(K,q)\propto\delta_{mn}e^{-(k_1^2+k_2^2)/2\sigma_k^2}$ with a possible
sign difference for the $bb$ component.
In Fig.~\ref{fig:Wavefunctions} we depict the spatial wavefunction $\Psi(X,\Delta)$ zooming in on the locations of the different bound states. 
All bound states show the typical exponential decay in the relative coordinate $\Delta$. The overall shapes of symmetric and anti-symmetric bound states are still close for small $\theta=\pi/10$, but there  are  larger 
deviations for 
$\theta=\pi/4$ where we have the maximum amplitude for tunneling into the $ab$ state. 
In this scenario the Hong-Ou-Mandel effect results in perfect switching between one photon in each waveguide ($ab$) and a  symmetric superposition of both photons in the same waveguide ($aa$ and $bb)$ after every single scattering event. At $\theta=\pi/4$ the symmetric bound state is the slowest with a delay time of $2/\Gamma$ per emitter since
it has the largest population in the $ab$ mode where the photons do not increase the speed of each other by blocking emitters.

While  the anti-symmetric  state $|B_K\rangle_{aa}-|B_K\rangle_{bb}$ is always a stable eigenstate state of the evolution due to symmetry, we have to resort to approximations to tackle the symmetric bound state for arbitrary $\theta$. For $K=0$ we observe that $\mathrm{S}|B_{K=0}\rangle=-|B_{K=0}\rangle$, while $\mathrm{S}_\mathrm{lin}|B_{K=0}\rangle=|B_{K=0}\rangle$ since $t_kt_{-k}=1$. This allows us to find a \emph{stable} eigenstate of the unit cell transformation \eqref{eq:unitTransformation} given by
\begin{align}\label{eq:SymmetricEigenstate}
|B_{K=0}\rangle_+\propto|B_{K=0}\rangle_{aa}+i\tan(\theta)|B_{K=0}\rangle_{ab} +|B_{K=0}\rangle_{bb},     
\end{align}
showing that for small tunneling probabilities the population in the $ab$ state  is negligible $\propto\theta^2$. 

For $K\neq 0$ the $ab$ component  picks up $k_1,k_2$ dependent phases relative to the $aa$ and $bb$ modes and we have not been able to identify exact eigenstates. We thus do not know if the state are true bound states or just metastable \cite{bakkensen2021photonic, calajo2022emergence,iversen2022self}. In the simulations, however,  we have a finite momentum width and thus probe the behavior around $K\approx 0$. The tunneling amplitude $\theta$ mainly affects the dispersion relation and we do not see strong differences between the bound state populations for the strongest tunneling $\theta=\pi/4$ and $\theta\rightarrow 0 $  (Fig.~\ref{fig:Wavefunctions} a and b) where $|B_K\rangle_{aa}+|B_K\rangle_{bb}$ is an exact eigenstate. This suggests that the symmetric state is a stable eigenstate for $K\approx 0 $.  
To test this we consider the extreme scenario of $\theta=\pi/4$ and 300 scatterings  (limited by our computational power). Even in this case  the symmetric bound state is hardly affected and thus stable for all practical purposes near $K=0$, see Supplements.

\emph{Conclusion--} 
We have shown that bound states made of photon-emitter polaritons act like truly composite particles when tunneling between waveguides or generally being transformed in beamsplitter-like fashion. Two photons of a bound state cannot tunnel individually between waveguides and especially for $\theta\ll1$ this results in a trapping of the bound state over periods of time where free polaritons would have already completely switched modes. Additionally, we identified the emergence of new bound state configurations in the setup of two close-by waveguides coupled to two-level emitters. Due to varying group velocities the different species of bound states can be spatially separated from each other as well as from the free polaritons. The Hong-Ou-Mandel effect at $\theta=\pi/4$, i.e.\,a 50/50 beamsplitter transformation between each pair of emitters, even allows for bound states with persistent self-oscillations. 
On a conceptual level, our results demonstrate that photon bound states show characteristics one would intuitively expect from a rigid molecule formed by photons. In the still mostly unexplored field of photons acting like bound matter, this promises for even more interesting phenomena, e.g. bound states interacting among themselves or being manipulated externally.

\emph{Acknowledgements--}  
B.\,S.\,is supported by Deutsche Forschungsgemeinschaft (DFG, German Research Foundation), Grant No. 449674892. We acknowledge the support of Danmarks Grundforskningsfond (DNRF 139, Hy-Q Center for Hybrid
Quantum Networks).

 \clearpage
 \appendix
\onecolumngrid
\begin{center}
\large{\textbf{Photon bound states: Supplemental material}}
\end{center}
\vspace{10pt}

\begin{figure*}
  \centering
  \includegraphics[width=0.99\textwidth]{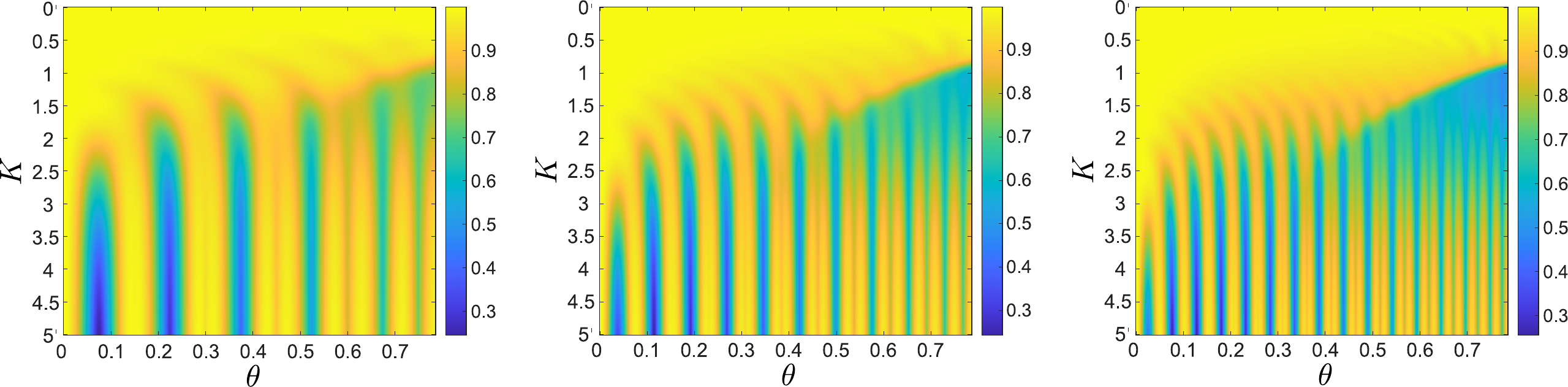}
  \caption{We study the stability of the symmetric bound state for $K\neq 0$ for different number of scatterings  $N=10$, $N=20$ and $N=30$ (left to right). The measure of stability is taken as the overlap with the exact eigenstate at $K = 0$  \eqref{eq:SymmetricEigenstate}. We observe a corridor of stability at $K \simeq 0$.}
\label{fig:Overlap}
\end{figure*}

In the main text we discuss the two newly emerging classes of bound states for coupled waveguides. While we can analytically show the stability of the anti-symmetric eigenstate, we can only describe the symmetric combination analytically for the special case of $K=0$. A broad numerical study suggests that for $K$ not close to 0 the symmetric state indeed decays, see Fig.~\ref{fig:Overlap}. For this numerical study we calculate the overlap after $10,20,30$ scatterings with the perfectly stable symmetric eigenstate \eqref{eq:SymmetricEigenstate} at $K=0$ for different tunneling strengths $\theta$ and center-of-mass momentum $K$. We observe that there is a broad region of stability for small $K$ that drops of sharply at large enough $K$. 

\begin{figure}[t!]
  \centering
  \includegraphics[width=0.99\textwidth]{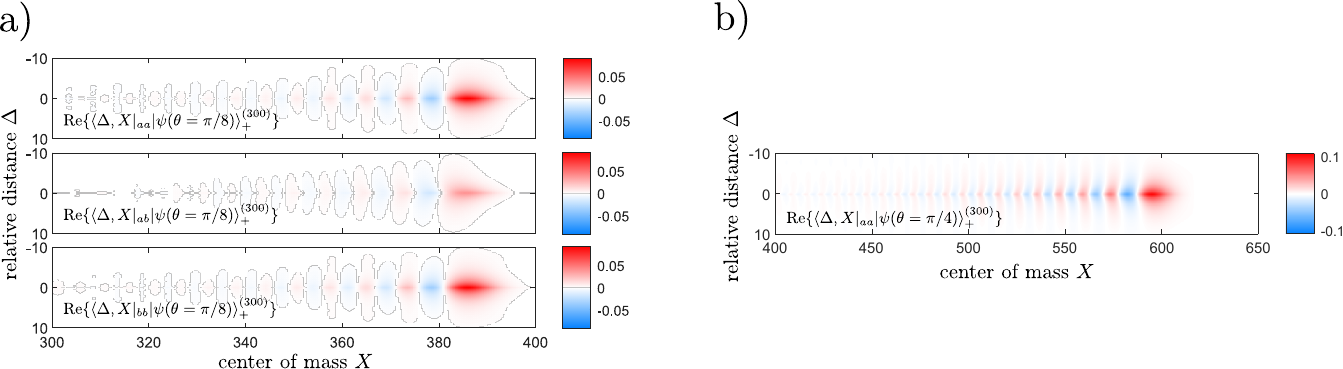}
  \caption{Wavefunction of the symmetric bound state at $\theta=\pi/8$ and $\theta=\pi/4$ after 300 scattering events. Because of the Hong-Ou-Mandel effect, in the latter case the state is zero in the $ab$ mode after the 300th scattering and the respective wave function is not shown.}
\label{fig:Wavefunctions300}
\end{figure}

This renders an input state tailored to maximize the bound state amplitude, e.g.\,a Gaussian with width $\sigma_k=\Gamma/3$, perfectly stable for all practical purposes. We show this exemplary for $\theta = \pi/8,\pi/4$ and 300 scattering events in Fig.~\ref{fig:Wavefunctions300}. Apart from ever present dispersion, the symmetric input state retains its bound state nature. We note that this numerical study is not a  proof of instability beyond certain $K$ as there is a possibility of  more complex eigenstate different from  \eqref{eq:SymmetricEigenstate} at finite $K$. Nevertheless, our findings do suggest that Eq.  \eqref{eq:SymmetricEigenstate} is an excellent approximation for a stable eigenstate at $K\simeq 0$.  

\begin{figure}
  \centering
  \includegraphics[width=0.60\textwidth]{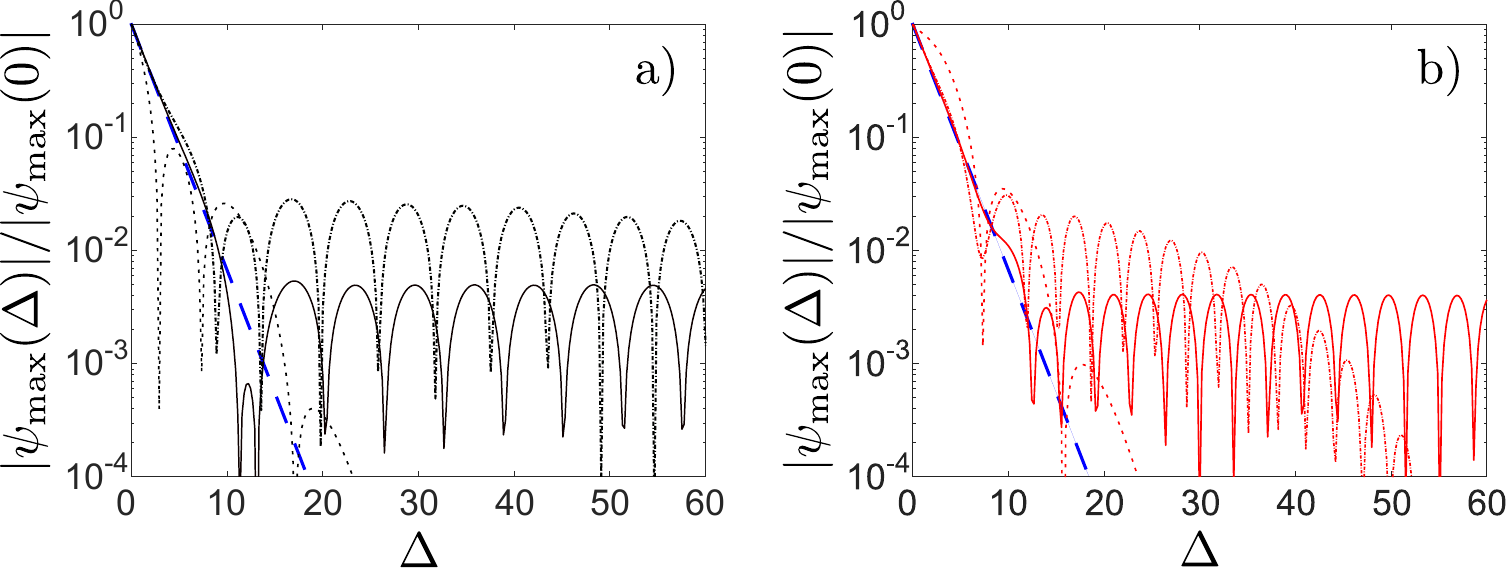}
  \caption{(a) Absolute value $|\psi_\mathrm{max}(\Delta)|$ of the real space wave function of the symmetric bound state at $\theta=\pi/4$, maximized over the center of mass coordinate $X$. The different curves depict the state after 3 (dotted lines), 30 (dashed-dotted lines), and 300 (solid lines) scatterings, all normalized to the maximum value. The dashed curve marks the ideal $e^{-\Gamma|\Delta|/2}$ decay of the bound state \eqref{eq:boundstate}. (b) Same as in (a) but for the anti-symmetric bound state as reference.}
\label{fig:LogTails}
\end{figure}

For completeness we also verify the $e^{-\Gamma|\Delta|/2}$ decay expected for bound states for both the symmetric and anti-symmetric input state, see Fig.~\ref{fig:LogTails}. Again inserting a Gaussian with $\sigma_k=\Gamma/3$ there is no real quantitative difference between the symmetric and anti-symmetric input state apart from oscillating corrections due to the unbound portion.

\end{document}